\documentclass[conference]{IEEEtran}
\IEEEoverridecommandlockouts

\usepackage{tabularx}
\usepackage{amsmath,amssymb,amsfonts}
\usepackage{algorithmic}
\usepackage{graphicx}
\usepackage{textcomp}
\usepackage{xcolor}
\usepackage[numbers,sort&compress]{natbib}
\usepackage{array}
\usepackage{amsmath}        
\usepackage[linesnumbered,ruled,vlined]{algorithm2e} 
\usepackage{booktabs}   
\usepackage{multirow}   
\usepackage{graphicx}   
\usepackage[T1]{fontenc}
\usepackage[hidelinks]{hyperref}

\usepackage{amssymb}  
\usepackage{array}    
\usepackage{pdfpages}
\usepackage{enumitem}
\usepackage{graphicx}
\usepackage{tcolorbox}
\usepackage{xcolor}
\usepackage{caption}
\definecolor{codeblue}{RGB}{0, 0, 255}      
\definecolor{codeteal}{RGB}{0, 128, 128}    
\definecolor{codegray}{RGB}{128, 128, 128}  
\tcbuselibrary{breakable, skins}

\newcommand{\xmark}{\textcolor{black!50}{--}}

\def\BibTeX{{\rm B\kern-.05em{\sc i\kern-.025em b}\kern-.08em
    T\kern-.1667em\lower.7ex\hbox{E}\kern-.125emX}}
\begin{document}

\title{MineValiCoder: Reliable Code Generation with Test Case Quality Mining and Bipartite Graph-Based Mutual Validation}

\author{
\IEEEauthorblockN{
\textbf{Zhen Zhao}\textsuperscript{1},
\textbf{Qihang Yang}\textsuperscript{2},
\textbf{Feifei Dai}\textsuperscript{1,*},
\textbf{Xiangfang Li}\textsuperscript{1},
\textbf{Bo Li}\textsuperscript{1}
}
\IEEEauthorblockA{
\textsuperscript{1}
Institute of Information Engineering, Chinese Academy of Sciences, Beijing, China\\
\texttt{\{zhaozhen, daifeifei, lixiangfang, libo\}@iie.ac.cn}
}
\IEEEauthorblockA{
\textsuperscript{2}
University of Electronic Science and Technology of China, Chengdu, China\\
\texttt{qihang.yang@std.uestc.edu.cn}
}
\thanks{* Corresponding author: Feifei Dai, e-mail: daifeifei@iie.ac.cn.}
}

\maketitle

\begin{abstract}
Large Language Model (LLM)-based Test-Driven Development (TDD) has advanced automated code generation. 
However, existing approaches depend heavily on human-crafted test cases and cannot operate with only natural-language requirements. 
Although recent works enable automatic test generation, they overlook the inherent stochasticity of LLMs, leading to two key defects: faulty tests generate misleading feedback that distorts code optimization, while mixed-quality test cases generate conflicting evaluation signals that hinder optimal code selection. 
To tackle these challenges, we rethink conventional TDD paradigms and propose that automated mutual reinforcement between test-case quality and code quality is essential for reliable, fully automated TDD. 
This paper presents MineValiCoder, a collaborative closed-loop TDD framework with three dedicated modules. 
The Test Case Quality Mining (TCQM) module filters out faulty test cases through self-validation mining, providing more reliable optimization supervision.
The Parallel TDD Refinement module iteratively optimizes code and generates diverse high-quality code candidates based on valid test-case feedback. 
The Bipartite Graph-Based Code-Test Mutual Validation (BiCoTeV) module dynamically models code-test interactions and performs mutual validation scoring to achieve stable and reliable selection of the optimal code. 
Extensive evaluations on four LLMs across mainstream benchmarks show that MineValiCoder significantly outperforms state-of-the-art methods.
Specifically, MineValiCoder achieves Pass@1 scores of 96.34\% on HumanEval, 87.40\% on MBPP, 64.00\% on APPS, and 51.33\% on LiveCodeBench.
These results strongly validate the effectiveness of our method in mitigating LLM stochasticity and substantially boosting the overall performance of automated code generation. Our code is available at: \url{https://anonymous.4open.science/r/MVCoder-E3DD}.
\end{abstract}


\section{Introduction}

Large Language Models (LLMs) have achieved competitive performance across a broad range of code generation tasks, ranging from lightweight function completion to full program synthesis \citep{roziere2023code,wang2023codet5+,achiam2023gpt,touvron2023llama}. 
To improve the reliability of generated code, substantial research has developed LLM-based Test-Driven Development (TDD) methods, which exploit execution feedback from test cases to iteratively refine LLM-generated code. 
For instance, Mathews and Nagappan~\citep{mathews2024test} validate generated code using human-crafted test cases and utilize error logs to guide iterative code refinement. 
Adnan et al.~\citep{adnan2025large} further present a self-debugging module that iteratively refines generated code using test execution results.
Despite promising results, these methods are inherently limited by their dependence on human-crafted test cases from existing datasets.
They fail to work in real-world scenarios where only natural-language requirements are available, leaving a critical problem to be addressed.

\begin{figure}[t]
\includegraphics[width=\columnwidth]{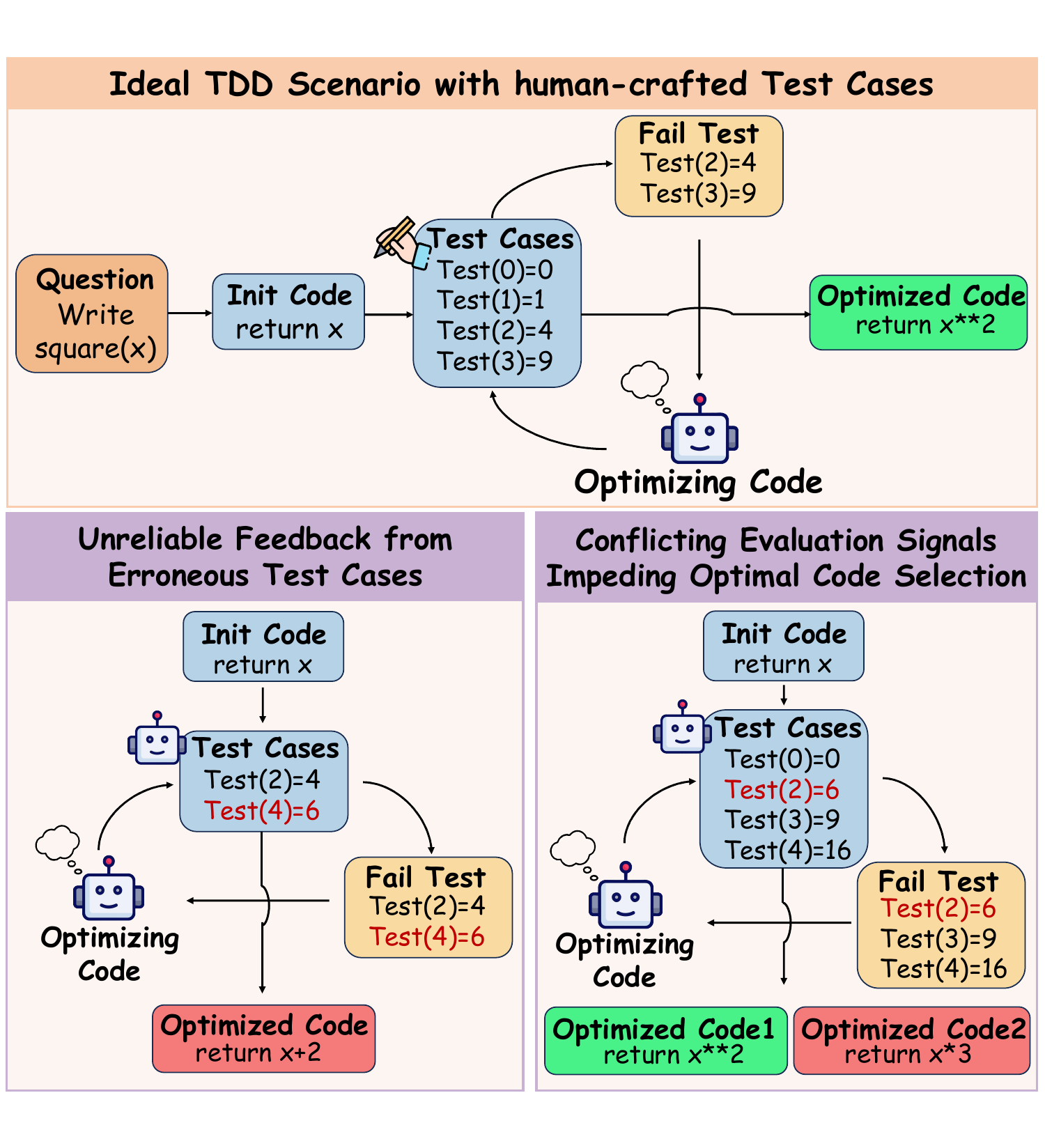}
\caption{Key challenges in automated TDD with LLM-generated test cases. Top: The ideal scenario guided by accurate human-crafted test cases. Bottom Left: Erroneous test cases provide unreliable feedback, misleading code optimization. Bottom Right: mixed-quality candidates make it difficult to select the optimal code.}
\label{fig:intro}
\end{figure}

To mitigate the reliance on human-crafted test cases and support TDD deployment with only natural-language requirements, recent studies turn to LLMs for automated test case generation. 
For example, Li and Yuan~\citep{li2024large} propose a multi-agent framework combined with a Python interpreter to generate test inputs and corresponding outputs, enabling the autonomous construction of complete test cases. 
Abdullin et al.~\citep{abdullin2025test} show that LLMs can comprehend code semantics and generate test cases with strong fault-detection capabilities.
Overall, these explorations lay a solid foundation for building fully automated TDD pipelines.

Nevertheless, simply combining the above techniques still presents key challenges in practice. 
Due to their inherent probabilistic characteristics, LLMs tend to produce imperfect test suites containing both valid and faulty test cases. 
Such noisy test cases severely impair the entire TDD workflow in two critical aspects:
\begin{itemize}
    \item \textbf{Faulty test cases lead to unreliable feedback that misleads code optimization.} 
    As shown in the bottom-left of Figure \ref{fig:intro}, when aiming to generate a $square(x)$ function, the generated test cases include an incorrect test case $\text{Test}(4)=6$. 
    Execution feedback based on this flawed test case misdirects the optimization process, causing the model to produce invalid code (\textit{i.e.}, \texttt{return x+2}) that overfits to the erroneous test case, as observed in prior work \citep{taherkhani2024valtest,wang2025codecontests+}. 

    \item \textbf{Mixed-quality test cases generate conflicting evaluation signals, hindering optimal code selection.} 
    As depicted in the bottom-right of Figure \ref{fig:intro}, the test cases include both valid cases ($\text{Test}(0)=0$, $\text{Test}(3)=9$, $\text{Test}(4)=16$) and an erroneous one ($\text{Test}(2)=6$). 
    The optimization process subsequently yields different candidates: the correct code \texttt{return x**2} and the flawed code \texttt{return x*3}. 
    Since both candidates pass different subsets of test cases (\texttt{x**2} passes valid cases; \texttt{x*3} passes the erroneous case and some valid ones), distinguishing the truly reliable code becomes extremely difficult.
\end{itemize}

To overcome the above two critical challenges, we re-examine the essence of automated TDD and propose a new design philosophy. 
Unlike conventional pipelines that treat test cases as fixed auxiliary inputs and focus only on optimizing code, we recognize that test-case quality and code quality are mutually reinforcing, and their intrinsic connections are key to resolving existing bottlenecks.
First, high-quality test cases serve as trustworthy supervision signals throughout iteration. 
By proactively filtering out faulty test cases upfront, we can deliver accurate execution feedback to guide code refinement. 
This fundamentally mitigates the risk of being misled by erroneous test cases, directly tackling the first challenge. 
Second, well-optimized code can in turn further verify and consolidate the reliability of test cases, forming a virtuous cycle of quality improvement for both sides.
Instead of judging code candidates merely by simple pass-or-fail results, we fully exploit the correlation between code and test cases for joint evaluation. 
This joint analysis can effectively reduce conflicting evaluation signals introduced by mixed-quality test cases, enabling us to accurately screen out high-reliability code from numerous candidates. This strategy directly addresses the second dilemma.

\begin{figure*}[]
    \centering
    \includegraphics[width=\textwidth]{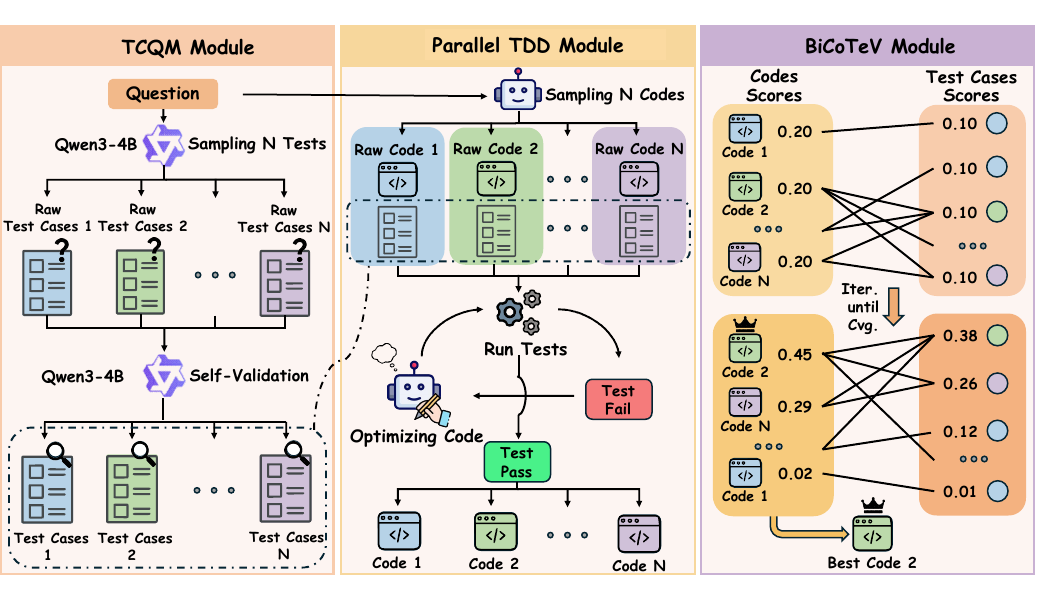}
    \caption{Overview of MineValiCoder. The framework first employs TCQM to generate validated test cases, followed by Parallel TDD for iterative code optimization. Finally, BiCoTeV leverages a bipartite graph to mutually validate code and test cases, mitigating LLM stochasticity to identify the optimal code candidate.}
    \label{fig:framework}
\end{figure*}

To materialize the aforementioned bidirectional quality enhancement and joint evaluation insights for resolving the two critical TDD bottlenecks, we propose a closed-loop automated TDD framework named MineValiCoder (Mining-Validation Enhanced Test-Driven Code Generation). MineValiCoder integrates automated test case quality mining and bipartite graph-based mutual validation to generate high-quality code. 
As illustrated in Figure \ref{fig:framework}, the framework comprises three core modules: 

(1) \textbf{Test Case Quality Mining (TCQM) Module.} To eliminate the dependency on human priors and prevent faulty test cases from yielding misleading feedback, this module first leverages LLMs to generate diverse batches of test cases from natural-language requirements, covering both common and edge-case scenarios.  
It further incorporates a self-validation mining mechanism to examine the input validity and logical rationality of each generated test case, filtering out faulty and low-quality test cases. 
In this way, only reliable test cases are retained to initialize the subsequent TDD workflow, mitigating the risk of misguided code refinement caused by flawed test cases.

(2) \textbf{Parallel TDD Refinement Module. } Different from conventional TDD pipelines that produce only a single final code version, this module performs parallel iterative refinement driven by the trustworthy test cases obtained from TCQM. It enables the LLM to explore multiple feasible implementation paths simultaneously and continuously corrects logical defects according to valid test feedback. By iteratively optimizing code under reliable test supervision, this module effectively mitigates optimization deviation caused by faulty test cases and produces a diverse set of high-quality code candidates with distinct implementation logic. Such diverse candidate pools provide a solid and comprehensive basis for the subsequent mutual validation and optimal selection in the BiCoTeV module.

(3) \textbf{Bipartite Graph-Based Code-Test Mutual Validation (BiCoTeV) Module.}  To address the conflicting evaluation signals caused by residual test noise and achieve accurate optimal code selection, this module models the relationship between code candidates and high-quality test cases as a bipartite graph and conducts dynamic mutual validation scoring. 
Code scores are updated based on the number of high-quality test cases they pass, while test case scores are adjusted by the consistency of code execution results (e.g., test cases passed by most valid code get higher scores). 
Ultimately, by ranking the candidates based on their scores, the algorithm identifies the optimal candidate code. 
By fully exploiting the inherent correlation between code and test cases, this module effectively alleviates the interference of noisy tests and enhances the stability of code selection.

Our contributions are summarized as follows:
\begin{itemize}
    \item We propose MineValiCoder, a fully automated closed-loop TDD framework that eliminates human prior dependence and enhances code quality amidst the inherent stochasticity of LLMs. By combining test quality mining and graph-based mutual validation, our framework addresses the limitations of existing LLM-powered TDD pipelines. 
    
    \item We design three complementary modules to address the key challenges in automated TDD. The Test Case Quality Mining (TCQM) Module automatically generates and validates test cases to filter out faulty ones, thereby mitigating misleading feedback during code optimization. The Parallel TDD Refinement Module performs stable iterative optimization across multiple independent branches to generate diverse code candidates. The Bipartite Graph-Based Code-Test Mutual Validation (BiCoTeV) Module models interactions between code candidates and test cases as a bipartite graph and adopts dynamic mutual validation scoring to reliably select the optimal code from multiple candidates.
    
    \item We conduct extensive experiments on four LLMs with different scales, including GPT-4, Llama3.1-8B, Qwen2.5-Coder-7B and Qwen3-4B, over several widely adopted benchmarks: HumanEval~\citep{chen2021evaluating}, MBPP~\citep{austin2021program}, APPS~\citep{hendrycks2021measuring} and LiveCodeBench~\citep{jain2025livecodebench}.
    Experimental results show that MineValiCoder achieves substantial improvements in code generation performance, with Pass@1 scores of $96.34\%$ on HumanEval, $87.40\%$ on MBPP, $64.00\%$ on APPS, and $51.33\%$ on LiveCodeBench. 
\end{itemize}

\section{Related Work}
Our work builds upon and extends two main lines of research in LLM-based code generation: 1) Direct Code Generation with LLMs, and 2) Iterative Code Refinement and Optimization with LLMs. This section reviews representative methods and highlights the specific gaps that our approach aims to address.

\subsection{Direct Code Generation with LLMs}
The impressive progress in code generation is largely driven by LLMs pre-trained on massive datasets of code and natural language~\citep{chen2021evaluating, achiam2023gpt, roziere2023code, li2023starcoder}. 
For example, CodeLlama~\citep{roziere2023code} is further trained on code-centric corpora to strengthen its programming ability, while StarCoder~\citep{li2023starcoder} is trained on large-scale permissively licensed source code covering diverse programming languages.
Given a natural-language problem description, these models can directly produce plausible and syntactically correct code, supporting a wide range of tasks from code completion to program synthesis.
However, due to the large and complex space of possible programs and the stochastic nature of LLM generation, the functional correctness of directly generated code is difficult to ensure.
Generated code may contain subtle logical errors or fail to handle edge cases, limiting its reliability in complex scenarios.
This fundamental limitation has motivated a shift towards iterative feedback-driven code refinement approaches.

\subsection{Iterative Code Refinement and Optimization with LLMs}
To overcome the limitations of direct code generation, recent studies have explored iterative refinement frameworks that use feedback, planning, or multi-agent collaboration to improve LLM-generated code.
Instead of producing code in a single attempt, these methods repeatedly evaluate, revise, and optimize candidate solutions, aiming to improve both functional correctness and overall code quality.
Existing approaches can be broadly characterized from two perspectives: the type of feedback used to guide refinement, and the way the refinement process is organized.

\textbf{Feedback-Guided Refinement.} 
As demonstrated by recent studies, the limitations of direct generation have driven the adoption of various forms of feedback to optimize the generated code. 
For instance, Self-Refine~\citep{madaan2023self} establishes a feedback-refinement loop in which LLMs produce textual self-criticism to drive iterative refinement of generated code. 
However, natural language feedback is often too coarse-grained to capture subtle boundary conditions and logical loopholes. 
To address this, Test-Driven Development (TDD) methods utilize unit test execution results as fine-grained feedback to optimize the code. 
For example, similar to PyCapsule~\citep{adnan2025large} and the approach of Mathews and Nagappan~\citep{mathews2024test}, BlueprintCode~\citep{mao2025blueprint2code} utilizes a multi-agent framework featuring a dedicated Debugging Agent that iteratively analyzes and repairs generated code based on execution feedback from dataset test cases. 
However, these TDD methods rely on human-crafted test cases, which are not always available.

\textbf{Organization of the Refinement Process.} 
To enhance the efficacy of iteration, recent works introduce more sophisticated mechanisms, primarily through advanced planning and multi-agent collaboration. 
One direction employs explicit planning and search, such as agent-guided tree search to systematically explore code candidate paths~\citep{li2025codetree}.
Another direction leverages multi-agent collaboration, designing frameworks where specialized agents (e.g., for planning, coding, testing, debating) interact to decompose problems, generate diverse candidates, or reach consensus. 
For instance, \citet{chen2025debatecoder} proposes a test-case-driven debate among LLM agents. \citet{islam2024mapcoder} employs specialized agents for planning, coding, and debugging. \citet{chen2024divide} decomposes problems via a divide-and-conquer strategy with consensus voting. 
These multi-agent approaches aim to improve code accuracy by achieving complex reasoning through task division and agent collaboration. 
However, the main limitation of such methods is that the complex collaborative reasoning mechanisms force the LLM to perform a long, multi-step intermediate reasoning process. 
During this process, due to the probabilistic nature of LLMs, the outputs involve a certain degree of randomness. 
Therefore, the final result cannot consistently yield the optimal code.

\paragraph*{Positioning of Our Work}
Our proposed \textbf{MineValiCoder} framework addresses the above limitations from two complementary perspectives.
First, it reduces the reliance on human-crafted test cases by automatically generating and mining high-quality test cases from natural-language requirements.
Second, it introduces a bipartite graph-based code-test mutual validation mechanism to improve code selection under residual test noise and stochastic LLM generation.
By integrating automated test case quality mining, iterative TDD refinement, and robust candidate selection, MineValiCoder provides a closed-loop framework for test-driven code generation.
This design enables more reliable code generation in settings where human-crafted test cases are unavailable or incomplete.

\section{Methodology}
\label{sec:methodology}
\subsection{Notations and Problem Definition}
In this paper, we focus on the task of code generation using Large Language Models (LLMs). Formally, given a natural language problem description $Q$, our objective is to utilize an LLM $M$ to generate the optimal code $c^*$. This process involves generating intermediate test cases $\mathcal{T}$ to guide the optimization of code.

Table~\ref{tab:notations} summarizes the necessary notations used throughout this paper.

\begin{table}[t]
	\centering
	\caption{Summary of Notations.}
	\label{tab:notations}
	\small 
	\begin{tabularx}{\columnwidth}{l >{\raggedright\arraybackslash}X}
		\toprule
		\textbf{Notation} & \textbf{Description} \\
		\midrule
		\multicolumn{2}{l}{\textit{\textbf{Problem Definition}}} \\
		$M$ & The Large Language Model (LLM) \\
		$Q$ & The natural language problem description \\
		$c^*$ & The final optimal code(Optimal Element) \\
		\midrule
		\multicolumn{2}{l}{\textit{\textbf{Test Case and Test Suite}}} \\
		$t$ & The test case is a pair of input ($I$) and expected output ($O$), where $t=(I, O)$ \\
		$T_i$ & The $i$-th test suite generated in round $i$, where $T_i=\{t_i^1,t_i^2,\ldots,t_i^{|T_i|}\}$ \\
		$\mathcal{T}$ & The collection of test cases from $N$ parallel rounds, where $\mathcal{T} = \bigcup\limits_{i=1}^{N} T_i$\\           
		\midrule
		\multicolumn{2}{l}{\textit{\textbf{Code}}} \\
		$\mathcal{C}$ & The set of code candidates \\
		$c_i$ & The $i$-th code candidate optimized specifically by suite $T_i$ \\
		\midrule
		\multicolumn{2}{l}{\textit{\textbf{Graph Validation}}} \\
		$E$ & The set of edges representing that a code $c$ passes a test case $t$ \\
		$\mathcal{G}$ & The bipartite graph composed of $\mathcal{C}$, $\mathcal{T}_{unique}$, and $E$ \\
		$Score(\cdot)$ & The reliability score of code candidates and test cases \\
		\bottomrule
	\end{tabularx}
\end{table}

\subsection{Overview}
To effectively address the aforementioned challenges, our framework \textbf{MineValiCoder} is designed around the core idea of \textit{optimizing and selecting code via test cases}. \textbf{MineValiCoder} comprises three modules, following a logical flow from test case generation to code optimization and final selection.

\begin{enumerate}[label=(\arabic*), leftmargin=*]
    
    
    \item \textbf{Test Case Quality Mining (TCQM)}:
    Since test cases serve as the foundation for the entire framework, their quality is critical to the final results.
    Therefore, this module aims to automate the mining of high-quality test cases.
    We first guide the LLM to generate diverse \textit{test suites} through prompt engineering.
    Specifically, we sample the LLM $N$ times to construct a set of $N$ independent test suites, denoted as $\{T_1,\dots,T_N\}$.
    Subsequently, at the \textit{test-case} level, we leverage the self-consistency capability of LLMs to filter out erroneous test cases within each suite.
    By designing a self-validation mining mechanism, we improve the accuracy of the retained test suites, thereby providing reliable optimization signals and high-quality test suites for subsequent steps.

    \item \textbf{Parallel TDD Refinement}: 
    To effectively use the mined test suites to optimize code, we design the \textbf{Parallel TDD Refinement} module. 
    First, we extend existing TDD approaches by introducing monotonicity-constrained optimization. 
    This makes the optimization process more stable and robust when handling LLM-generated test suites. 
    Secondly, to increase the probability of obtaining an optimal code, this module utilizes the mined test suites $\{T_1,\dots,T_N\}$ to drive $N$ parallel TDD pipelines.
    Each pipeline independently optimizes the code based on feedback from its corresponding test suite, yielding a diverse set of high-quality code candidates $\mathcal{C} = \{c_1, \dots, c_N\}$.

    \item \textbf{Bipartite Graph-Based Code-Test Mutual Validation (BiCoTeV)}: 
    To address the conflicting evaluation signals caused by residual test noise and select the optimal code from generated code candidates, we design the \textbf{BiCoTeV} module. 
    In this module, we model the relationship between the code candidates and the global test cases as a bipartite graph. 
    The code candidates $\mathcal{C}$ and the global test cases $\mathcal{T}$ serve as the two disjoint sets of nodes, while the execution results (i.e., pass/fail status) constitute the edges connecting them. 
    Based on this graph construction, we employ a dynamic score-propagation algorithm to perform global mutual validation. 
    By fully exploiting the mutual-validation relationships between code candidates and global test cases, this module allows us to select the optimal code $c^*$ based on the final score.

\end{enumerate}

\subsection{Test Case Quality Mining (TCQM)}
\label{subsec:tcqm}

A test case $t$ consists of an input $I$ and an expected output $O$.
To validate code correctness, we run the candidate code on $I$ and compare its execution output with $O$.
This comparison yields a boolean result (pass/fail) that reflects the code's functional accuracy. 
A complete test suite $T$ is a collection of multiple test cases. 
To fully examine code behavior, a qualified test suite must cover diverse input scenarios.
Accordingly, the overall quality of a test suite depends not only on the correctness of individual test cases but also on its scenario diversity. 
Aiming to address the issue of faulty test cases generated by LLMs, and to mitigate misleading feedback in subsequent code optimization, the TCQM module refines raw test case outputs in two sequential stages:

\textbf{Diversity Generation.} 
To generate diverse test cases, we design a prompt $P_{diverse}$ that guides the LLM to cover a wide range of input scenarios automatically.
Specifically, the generation prompt explicitly mandates the inclusion of \textit{normal scenarios}, \textit{boundary conditions}, and \textit{problem-specific test cases}.
Aligned with the framework's parallel architecture, we employ $N$ rounds of independent sampling in this stage.
The entire generation process can be formulated as:

\begin{equation}
	\hat{T}_i = M(P_{diverse}(Q)), \quad \text{for } i = 1, \dots, N,
\end{equation}

where $\hat{T}_i$ denotes the raw test suite generated in the $i$-th pipeline based on the problem description $Q$ and the prompt $P_{diverse}$.

\textbf{Self-Validation.} 
Due to the inherent stochasticity of LLMs, raw LLM-generated test cases often contain logical errors and hallucinations. 
Inspired by Self-Consistency ~\citep{wangself}, which observes that valid reasoning tends to yield consistent results across multiple independent runs, we adopt this property to filter out defective test cases.
We construct a validation prompt $P_{validation}$ to re-infer the expected output for each test case. 
Formally, for each generated test case $t=(I, O) \in \hat{T}_i$, we mask the original output $O$ and prompt the model to predict the output $\hat{O}$ given input $I$ and problem $Q$. 
The validated test suite $T_i$ is constructed by retaining only test cases with consistent outputs:

\begin{equation}
\begin{split}
    T_i = \bigl\{ (I, O) \in \hat{T}_i \mid \hat{O} = O, \\
    \text{where } \hat{O} \leftarrow M(P_{validation}(Q, I)) \bigr\}.
\end{split}
\end{equation}

This rigorous validation ensures that the entire TDD workflow is driven by reliable test cases, substantially reducing the risk that code optimization is misled by flawed tests.

\subsection{Parallel TDD Refinement}
\label{subsec:tdd_refinement}

Building upon the high-quality test suites $\{T_1,\dots,T_N\}$ constructed in the TCQM module, this module aims to generate a diverse set of high-quality and robust code candidates $C$.
Different from prior TDD approaches that blindly optimize toward all failed test cases, this module combines parallel iterative optimization with monotonicity constraints.

Running $N$ fully independent optimization branches enables diverse exploration of the solution space, which reduces the risk of becoming trapped in local optima inherent in single-path repair and greatly enriches the heterogeneity of the final code candidate pool. 
To further guarantee the optimization quality of each parallel branch, we integrate monotonicity optimization into the parallel TDD pipeline to standardize every code repair iteration. 
Specifically, each newly optimized code is required to fix target failed test cases without invalidating any previously passed valid test cases. 
This lightweight yet effective constraint substantially suppresses the negative interference of residual erroneous test cases, ensuring stable and sustainable quality improvement for all parallel optimization paths. 
Finally, these independent parallel TDD pipelines produce a high-quality and diverse code candidate pool, laying a solid foundation for the global mutual validation of the downstream BiCoTeV module. 

Algorithm~\ref{alg:tdd_process} describes the core iterative TDD process, which is designed to ensure stable and effective code optimization.
Each pipeline employs a specific test suite $T_i$ to provide fine-grained feedback for iteratively refining the initial code through the following steps:

(1) Initialization and Partitioning. To identify test cases for optimization, the process initializes the code candidate $c_{curr}$ and immediately executes it against the test suite $T_i$. 
This execution partitions the test cases into passed ($\mathcal{P}$) and failed ($\mathcal{F}$) sets, where $\mathcal{F}$ represents the cases needed to be optimized (lines 1--2). 
Additionally, a set $\mathcal{U}$ is initialized to track unfixable test cases (line 3).

(2) Iterative Repair Loop. 
To repair as many failed test cases as possible, the main optimization loop (lines 4--21) persists as long as there exists a failing test case $f \in \mathcal{F}$ that has not yet been marked as unfixable. 
Inside this loop, the function attempts to repair the target $f$ up to $R_{max}$ times (lines 6--17).

(3) "Plan-then-Code" Generation. To improve code quality, lines 7--8 employ a "Plan-then-Code" mechanism. Instead of generating code directly—which often leads to short-sighted patches—the LLM first formulates an optimization plan based on the feedback of $f$, and then generates the updated code $c_{new}$. This decomposition forces the LLM to reason about the root cause before implementation.

(4) Monotonicity Optimization. To avoid code quality degradation caused by overfitting, lines 10--16 enforce a strict monotonicity constraint. Line 9 re-executes $c_{new}$ on $T_i$ to obtain $\mathcal{P}_{new}$ and $\mathcal{F}_{new}$. 
$c_{new}$ is accepted if and only if it fixes the target $f$ without failing on any previously passed cases in $\mathcal{P}$. 
This guarantees that the optimization proceeds in a strictly ascending direction, preventing the "fix one, break two" issue.

(5) Unfixable Handling. Finally, if $f$ remains unfixed after the maximum number of retries $R_{max}$, it is added to the $\mathcal{U}$ set (lines 18--20). This strategy prevents the model from entering infinite loops on erroneous test cases.
\begin{algorithm}[t]
\caption{Iterative TDD Refinement}
\label{alg:tdd_process}
\SetAlgoLined
\KwIn{Problem $Q$, Test Suite $T_i$, Max Retries $R_{max}$}
\KwOut{Optimized Code Candidate $c_{curr}$}

$c_{curr} \leftarrow M(Q)$\;
$\mathcal{P}, \mathcal{F} \leftarrow \text{Execute}(c_{curr}, T_i)$\;
$\mathcal{U} \leftarrow \emptyset$\;

\While{$\exists f \in \mathcal{F}$ \textbf{such that} $f \notin \mathcal{U}$}{
    
    $fixed \leftarrow \textbf{false}$\;
    
    \For{$r \leftarrow 1$ \KwTo $R_{max}$}{
        $Plan \leftarrow M(Q, c_{curr}, f)$\;
        $c_{new} \leftarrow M(c_{curr}, Plan)$\;
        
        $\mathcal{P}_{new}, \mathcal{F}_{new} \leftarrow \text{Execute}(c_{new}, T_i)$\;
        
        \If{$f \in \mathcal{P}_{new}$ \textbf{and} $\mathcal{P} \subseteq \mathcal{P}_{new}$}{
            $c_{curr} \leftarrow c_{new}$\;
            $\mathcal{P} \leftarrow \mathcal{P}_{new}$\;
            $\mathcal{F} \leftarrow \mathcal{F}_{new}$\;
            $fixed \leftarrow \textbf{true}$\;
            \textbf{break}\; 
        }
    }
    
    \If{\textbf{not} $fixed$}{
        $\mathcal{U} \leftarrow \mathcal{U} \cup \{f\}$\;
    }
}
\Return $c_{curr}$\;
\end{algorithm}

After executing $N$ parallel and independent TDD refinement processes, we collect all optimized outputs to form a diverse and high-quality code candidate set $\mathcal{C} = \{c_1, c_2, \dots, c_N\}$, which is fed into the subsequent BiCoTeV module for optimal code selection.

\subsection{Bipartite Graph-based Code-Test Mutual Validation (BiCoTeV)}
\label{subsec:bicotev}
To identify the optimal code $c^*$ from the diverse code candidate set $\mathcal{C}$, we design the \textbf{BiCoTeV} module.

Despite the faulty-test filtering performed by TCQM, the inherent stochasticity of LLMs may still leave residual errors in the retained test cases. 
Such imperfect test cases can generate conflicting evaluation signals and introduce bias during candidate selection.
To address this issue, we adopt a mutual validation strategy to simultaneously identify high-quality code and test cases. 
Our core insight is as follows: reliable test cases tend to be passed by most high-quality code candidates, and excellent code can consistently pass tests with high credibility. 

Specifically, to identify high-quality code candidates and test cases simultaneously, we model code candidates $\mathcal{C}$ and global test cases $\mathcal{T}$ as a bipartite graph. In this structure, $\mathcal{C}$ and $\mathcal{T}$ serve as the two disjoint sets of nodes, while the execution results (i.e., pass/fail status) constitute the edges connecting them. To effectively evaluate the respective quality of $\mathcal{C}$ and $\mathcal{T}$, we employ a dynamic score-propagation algorithm to perform mutual validation. Specifically, reliability scores propagate bidirectionally along the edges: a candidate's score is aggregated from the scores of the tests it passes, while a test case's score is dynamically calculated based on the candidates it validates. This process converges to identify the optimal code $c^*$. The implementation comprises three sequential steps:

\textbf{Bipartite Graph Construction.} We construct a bipartite graph $\mathcal{G} = (\mathcal{C}, \mathcal{T}_{unique}, E)$. In this graph, the edges represent the validation relationship: an edge $(c_i, t_j) \in E$ is established if and only if the code candidate $c_i \in \mathcal{C}$ successfully passes the specific test case $t_j \in \mathcal{T}$.

\textbf{Dynamic Mutual Validation Scoring.} To reasonably evaluate the quality of code and test cases, we introduce a dynamic mutual validation scoring algorithm on the bipartite graph. Let $Score_k(\cdot)$ denote the score at the $k$-th iteration, and $d$ be the damping factor. We synchronously update the scores of both node types using the following formulas:
\begin{align}
    \begin{split}
        Score_k(c_i) &= (1 - d) \cdot Score_{k-1}(c_i) \\
                     &\quad + d \cdot \sum_{t_j \in \mathcal{T}_{unique}} Score_{k-1}(t_j) \cdot \mathbb{I}((c_i, t_j) \in E),
    \end{split} \\
    \begin{split}
        Score_k(t_j) &= (1 - d) \cdot Score_{k-1}(t_j) \\
                     &\quad + d \cdot \sum_{c_i \in \mathcal{C}} Score_{k-1}(c_i) \cdot \mathbb{I}((c_i, t_j) \in E),
    \end{split}
\end{align}
where $\mathbb{I}((c_i, t_j) \in E)$ is the indicator function that equals 1 if the edge exists (i.e., $c_i$ passes $t_j$), and 0 otherwise.  For numerical stability, scores are normalized such that $\sum S = 1$ after each iteration. This process iterates until convergence or the maximum number of iterations is reached. This mechanism effectively down-weights easy tests (passed by all) and penalizes codes that only pass unreliable tests.

\textbf{Consistency-based Robust Filtering.} 
While higher scores typically indicate superior code quality, they can also result from overfitting to a larger number of erroneous test cases.  
To address this, we design a consistency-based filtering mechanism. 
For each code candidate $c_i$, we analyze its set of passed test cases $\mathcal{T}_{pass}$, and its set of failed test cases $\mathcal{T}_{fail}$. 
A candidate is identified as overfitting and filtered out if the maximum score among the failed cases exceeds the minimum score among the passed cases: 
\begin{equation}
\max_{t \in \mathcal{T}_{fail}} Score(t) > \min_{t \in \mathcal{T}_{pass}} Score(t).
\end{equation}
This criterion reveals that the code performs well on less reliable test cases but fails on highly credible ones, which is a typical sign of overfitting. 
After filtering out problematic candidates, we select the one with the highest score from the remaining pool as the final optimal code $c^*$.

\begin{table*}[t]
\centering
\caption{Performance comparison of code generation methods across multiple benchmarks. We report Pass@1 (\%) for all tasks, where ``-ET'' denotes the enhanced variant of each dataset and ``-'' indicates unavailable results. Best results are highlighted in \textbf{bold}. For the APPS dataset, \textbf{Intro}, \textbf{Inter}, and \textbf{Comp} denote the \textit{Introductory}, \textit{Interview}, and \textit{Competition} difficulty levels, respectively, with \textbf{Avg} representing the overall average. Note that MapCoder~\citep{islam2024mapcoder} is incompatible with Llama3.1-8B and Qwen3-4B in our experiments.}
\label{tab:main_results}
\resizebox{\textwidth}{!}{
\begin{tabular}{llccccccccc}
\toprule
\multirow{2}{*}{\textbf{Model}} & \multirow{2}{*}{\textbf{Method}} & \multirow{2}{*}{\textbf{HumanEval}} & \multirow{2}{*}{\textbf{HumanEval-ET}} & \multirow{2}{*}{\textbf{MBPP}} & \multirow{2}{*}{\textbf{MBPP-ET}} & \multirow{2}{*}{\textbf{LiveCodeBench}} & \multicolumn{4}{c}{\textbf{APPS}} \\
\cmidrule(lr){8-11} 
 &  &  &  &  &  &  & \textbf{Intro} & \textbf{Inter} & \textbf{Comp} & \textbf{Avg} \\
\midrule

\multirow{7}{*}{GPT-4} 
 & Direct & 80.10 & 73.80 & 81.10 & 57.70 & 38.67 & 40.00 & 34.00 & 18.00 & 30.67 \\
 & CoT\citep{wei2022chain} & 89.00 & 61.60 & 82.40 & 56.20 & 42.67 & 52.00 & 52.00 & 26.00 & 43.33 \\
 & Self-Planning\citep{jiang2024self} & 85.40 & 62.20 & 75.80 & 50.40 & 48.00 & 58.00 & 56.00 & 40.00 & 51.33 \\
 & Reflexion~\citep{shinn2023reflexion} & 91.00 & 78.70 & 78.30 & 51.90 & - & - & - & - & - \\
 & CodeT~\citep{chencodet} & 92.07 & 84.15 & 86.50 & 68.10 & 41.33 & 54.00 & 50.00 & 32.00 & 45.33 \\
 & MapCoder\citep{islam2024mapcoder} & 93.90 & 82.90 & 83.10 & 57.70 & 48.00 & 62.00 & 50.00 & 28.00 & 46.67 \\
 & \textbf{MineValiCoder (Ours)} & \textbf{96.34} & \textbf{89.63} & \textbf{87.40} & \textbf{70.10} & \textbf{51.33} & \textbf{74.00} & \textbf{66.00} & \textbf{52.00} & \textbf{64.00} \\
\midrule

\multirow{5}{*}{\shortstack[l]{Qwen2.5-\\Coder-7B}} 
 & Direct & 87.80 & 81.71 & 82.90 & 67.00 & 16.67 & 1.00 & 9.00 & 2.00 & 4.00 \\
 & CoT\citep{wei2022chain} & 87.80 & 80.49 & 79.90 & 64.40 & 15.33 & 8.00 & 12.00 & 4.00 & 8.00 \\
 & CodeT~\citep{chencodet} & 90.85 & 84.76 & 87.40 & 69.50 & 19.33 & 16.00 & 12.00 & 2.00 & 10.00 \\
 & MapCoder\citep{islam2024mapcoder} & 85.37 & 78.05 & 88.40 & 71.10 & 24.67 & 16.00 & 10.00 & 2.00 & 9.33 \\
 & \textbf{MineValiCoder (Ours)} & \textbf{93.29} & \textbf{86.59} & \textbf{88.90} & \textbf{70.60} & \textbf{26.67} & \textbf{34.00} & \textbf{18.00} & \textbf{6.00} & \textbf{19.33} \\
\midrule

\multirow{5}{*}{\shortstack[l]{Qwen3-\\4B}} 
 & Direct & 76.22 & 71.34 & 73.80 & 58.70 & 21.33 & 18.00 & 16.00 & 8.00 & 14.00 \\
 & CoT\citep{wei2022chain} & 79.88 & 71.95 & 69.30 & 55.60 & 21.33 & 28.00 & 16.00 & 18.00 & 20.67 \\
 & CodeT~\citep{chencodet} & 82.31 & 75.61 & 78.80 & 62.30 & 22.00 & 28.00 & 26.00 & 16.00 & 23.33 \\
 & \textbf{MineValiCoder (Ours)} & \textbf{92.07} & \textbf{86.59} & \textbf{82.90} & \textbf{64.90} & \textbf{28.67} & \textbf{38.00} & \textbf{24.00} & \textbf{20.00} & \textbf{27.33} \\
\midrule

\multirow{5}{*}{\shortstack[l]{Llama3.1-\\8B}} 
 & Direct & 63.41 & 56.70 & 73.30 & 57.70 & 8.67 & 2.00 & 6.00 & \textbf{2.00} & 3.33 \\
 & CoT\citep{wei2022chain} & 60.97 & 54.27 & 61.30 & 48.90 & 14.00 & 10.00 & 4.00 & 0.00 & 4.67 \\
 & CodeT~\citep{chencodet} & 74.39 & 66.46 & 74.80 & 59.20 & 10.67 & 6.00 & 4.00 & \textbf{2.00} & 4.00 \\
 & \textbf{MineValiCoder (Ours)} & \textbf{84.76} & \textbf{75.60} & \textbf{81.40} & \textbf{63.90} & \textbf{18.00} & \textbf{14.00} & \textbf{10.00} & \textbf{2.00} & \textbf{8.67} \\
\bottomrule
\end{tabular}
}
\end{table*}

\section{Experiments}
\label{sec:experiments}

\subsection{Experimental Setup}
\label{subsec:setup}

\noindent \textbf{Datasets.} We evaluate \textbf{MineValiCoder} on six widely adopted code generation benchmarks, which cover tasks spanning from basic function implementation to challenging competitive programming problems. The evaluation encompasses two main categories of datasets: \textbf{Single-function algorithmic tasks} include HumanEval \citep{chen2021evaluating}, MBPP \citep{austin2021program}, as well as their enhanced variants HumanEval-ET and MBPP-ET \citep{liu2023your}. These benchmarks enable rigorous assessment of code generation quality; \textbf{Competition program generation tasks} provide complex, competition-level programming. We adopt a balanced subset of \textbf{APPS}~\citep{hendrycks2021measuring} with 150 problems, where 50 samples are selected from each of the \textit{Introductory}, \textit{Interview}, and \textit{Competition} difficulty tiers. To eliminate data leakage risks for newly released models, we additionally use 150 randomly sampled problems from \textbf{LiveCodeBench}~\citep{jain2025livecodebench}. For all datasets, we adopt \textbf{Pass@1} as the evaluation metric. A generated code solution is deemed correct if and only if it passes all corresponding test cases.

\textbf{Baselines.} 
To comprehensively evaluate the effectiveness of \textbf{MineValiCoder}, we select two main categories of mainstream methods as baselines. The first category consists of general LLM reasoning-enhancement methods applicable across different tasks, while the second consists of code-specific optimization frameworks tailored to code-generation problems.

\textbf{1. General LLM reasoning enhancement methods.} These methods enhance code generation by modifying the input prompt or reasoning process.
\begin{itemize}
    \item \textbf{Direct Prompting:} The model generates code directly from the problem description without intermediate reasoning.
    \item \textbf{Chain-of-Thought (CoT)} \citep{wei2022chain}: The model produces step-by-step reasoning before generating the final code solution.
    \item \textbf{Self-Planning} \citep{jiang2024self}: The task is decomposed into separate planning and implementation phases.
    \item \textbf{Reflexion} \citep{shinn2023reflexion}: The model iteratively optimizes code based on feedback from execution errors.
\end{itemize}

\textbf{2. Code-Specific Optimization Frameworks.} These approaches explicitly leverage program execution, test generation, or specialized architectures to iteratively produce or select correct code.
\begin{itemize}
    \item \textbf{CodeT} \citep{chencodet}: Multiple code candidates and corresponding test cases are generated, and majority voting over test-case execution results is used for final selection.
    \item \textbf{MapCoder} \citep{islam2024mapcoder}: A multi-agent system simulates the software development lifecycle through retrieval, planning, coding, and debugging agents.
    \item \textbf{MineValiCoder(Ours)}: Our framework mines high-quality test cases and code candidates, and performs joint validation and ranking to determine the optimal solution.
\end{itemize}

\textbf{Models.} To make our evaluation comprehensive and representative, we adopt a set of widely used open-source and closed-source LLMs as the backbone models. We select \textbf{GPT-4}~\citep{achiam2023gpt}, \textbf{Qwen2.5-Coder-7B}~\citep{hui2024qwen2}, \textbf{Llama3.1-8B}~\citep{grattafiori2024llama} and \textbf{Qwen3-4B}~\citep{yang2025qwen3} for experiments. Notably, we use Qwen3-4B~\citep{yang2025qwen3} as the dedicated backbone for test case generation in the TCQM module. This configuration verifies that our test case mining framework does not depend on the strong inherent capabilities of large models such as GPT-4~\citep{achiam2023gpt}. Instead, it leverages validation and consensus mechanisms to convert weak signals from lightweight models into reliable supervision, delivering promising performance with high efficiency.

\textbf{Implementation Details.} We set the sampling size $N=8$ to control the number of both mined test cases and code candidates for each problem. This hyperparameter was determined through systematic experiments to achieve an optimal tradeoff between accuracy and cost. To suppress the generation of unreasonable low-probability content, we apply nucleus sampling with \texttt{top\_p}=0.9 across all decoding processes. We adopt a relatively low temperature of 0.6 during code optimization to ensure logically stable refinements. For code generation, we employ varied temperatures of $[0.8, 1.2, 1.5]$ to encourage diversity and broader exploration of the code space. For the BiCoTeV module, iteration terminates when either the score difference between consecutive iterations falling below $10^{-5}$, or the total iteration number reaches 30. All experiments were conducted on two NVIDIA A100 GPUs.

\subsection{Results}
\label{subsec:results}

As shown in Table \ref{tab:main_results}, \textbf{MineValiCoder} outperforms all baselines and achieves state-of-the-art results when built upon GPT-4. This remarkable performance demonstrates the efficacy of our overall framework.

\textbf{Superior Performance on Closed-Source Model(GPT-4).} 
On single-function algorithmic benchmarks, \textbf{MineValiCoder} achieves state-of-the-art performance across all datasets. 
Specifically, it attains a Pass@1 score of 96.34\% on HumanEval, surpassing the previous best (MapCoder, 93.90\%). On HumanEval-ET, it scores 89.63\%, outperforming strong baselines like CodeT (84.15\%). 
The performance lead extends to MBPP with a score of 87.40\% (vs. CodeT's 86.50\%) and to MBPP-ET with 70.10\% (vs. CodeT's 68.10\%).

We further evaluate our approach on the challenging APPS benchmark, which requires sophisticated reasoning and algorithm design. 
MineValiCoder maintains the leading performance on all difficulty levels, with Pass@1 scores of 74.00\%, 66.00\% and 52.00\% for Introductory, Interview and Competition tasks, respectively. 
This comprehensive advantage leads to a new state-of-the-art average score of 64.00\%, which exceeds the previous best method (Self-Planning, 51.33\%) by a significant margin of +12.67\%. The results demonstrate that our design is particularly effective at mitigating stochastic errors in complex code generation.

Due to potential data leakage issues associated with LLMs, we conduct additional experiments on LiveCodeBench to thoroughly examine the real problem-solving capability of MineValiCoder. 
This benchmark consists of continuously updated competitive programming problems, which can effectively prevent models from relying on training data memorization. 
On this uncontaminated benchmark, \textbf{MineValiCoder} continues to achieve the highest performance with a Pass@1 score of 51.33\%, outperforming the strongest baselines, Self-Planning and MapCoder (both at 48.00\%). This consistent superiority strongly demonstrates that the performance gains of \textbf{MineValiCoder} stem directly from its inherent mechanism rather than dataset memorization. Consequently, this further validates its robustness and generalization ability in competition program generation tasks.

\textbf{Empowering Lightweight Open-Source Models.} Our method brings substantial and consistent performance improvements to lightweight open-source models, including  Qwen3-4B, Qwen2.5-Coder-7B and Llama3.1-8B, across all benchmarks. 
MineValiCoder achieves state-of-the-art results in all settings. 
These consistent gains verify the general effectiveness and strong applicability of our framework.

\begin{table}[htbp]
  \centering
  \caption{Performance comparison with PyCapsule. PyCapsule utilizes ground-truth tests from dataset, serving as a performance upper bound for TDD-method code generation. MineValiCoder achieves highly comparable results using solely LLM-generated tests.}
  \label{tab:upper-bound-comparison}
  \begin{tabular*}{\columnwidth}{@{\extracolsep{\fill}} llcc @{}}
    \toprule
    \textbf{Method} & \textbf{Test Source} & \textbf{HumanEval (\%)} & \textbf{MBPP (\%)} \\
    \midrule
    PyCapsule & Ground-truth & 96.50 & 88.20 \\
    \midrule
    \textbf{MineValiCoder} & \textbf{Self-generated} & \textbf{96.34} & \textbf{87.40} \\
    \bottomrule
  \end{tabular*}
\end{table}

\textbf{Comparison with Human-Crafted-Test-Based TDD Frameworks.} As shown in Table~\ref{tab:upper-bound-comparison}, we further compare MineValiCoder with PyCapsule. By utilizing ground-truth tests, PyCapsule establishes a performance upper bound for TDD-method code generation. PyCapsule reports Pass@1 of 96.50\% on HumanEval and 88.20\% on MBPP, whereas MineValiCoder reaches 96.34\% and 87.40\%, respectively, using only LLM-generated tests. These remarkably close results demonstrate that the accuracy of MineValiCoder closely approaches this reference upper bound, without relying on human-crafted test cases.

In summary, experimental results demonstrate that \textbf{MineValiCoder} achieves superior performance across all evaluated benchmarks and model scales. With GPT-4, it establishes new state-of-the-art results, notably achieving a 96.34\% Pass@1 on HumanEval and a remarkable 64.00\% average on the challenging APPS benchmark. The framework also democratizes high performance to open-source models (Qwen2.5-Coder-7B, Qwen3-4B, Llama3.1-8B), delivering consistent and substantial gains across all tasks.

\subsection{Ablation Study}

\label{subsec:ablation}
\paragraph{Ablation study of modules}
To assess the contribution of each component in \textbf{MineValiCoder}, we conduct ablation experiments on HumanEval using GPT-4 as the code-generation backbone. 
GPT-4 is adopted as a strong backbone among the evaluated models, making the ablation results less dependent on the limitations of weaker base models. 
We choose HumanEval because it provides standardized execution-based evaluation for clean single-function algorithmic tasks, which enables a clear and comparable analysis of individual modules. 
We do not use APPS or LiveCodeBench for this ablation because competetion-program tasks involve more diverse input-output formats, longer problem contexts, and larger difficulty variations, which may introduce additional confounding factors for module-level analysis. 
The broader effectiveness of our framework is further evaluated across all benchmarks in the main experiments.

\begin{table}[htbp]
\centering
\caption{Ablation study of different modules on the HumanEval benchmark, presenting the performance improvements when incorporating each component individually and in combination, where "\xmark" indicates the module is not included, "\checkmark" indicates the module is included, and "$\Delta$" denotes the improvement relative to the baseline.}
\label{tab:ablation_fullwidth}

\begin{tabular*}{\linewidth}{@{\extracolsep{\fill}} ccc cc}
\toprule
\multicolumn{3}{c}{\textbf{Components}} & \multicolumn{2}{c}{\textbf{Results}} \\
\cmidrule(lr){1-3} \cmidrule(lr){4-5} 

\textbf{TCQM} & \textbf{Parallel TDD} & \textbf{BiCoTeV} & \textbf{Pass@1 (\%)} & \textbf{$\Delta$} \\
\midrule
\xmark & \xmark & \xmark & 80.10 & - \\
\xmark & \checkmark & \xmark & 87.80 & +7.70 \\
\xmark & \xmark & \checkmark & 89.63 & +9.53 \\
\midrule
\xmark & \checkmark & \checkmark & 90.85 & +10.75 \\
\checkmark & \xmark & \checkmark & 93.90 & +13.80 \\
\checkmark & \checkmark & \xmark & 92.68 & +12.58 \\
\midrule
\textbf{\checkmark} & \textbf{\checkmark} & \textbf{\checkmark} & \textbf{96.34} & \textbf{+16.24} \\
\bottomrule
\end{tabular*}
\end{table}

The ablation results are presented in Table~\ref{tab:ablation_fullwidth}. Using the base GPT-4 model with direct prompting (all components disabled) achieves a Pass@1 score of 80.10\%, which serves as our baseline. 
Individually enabling the components reveals that \textbf{BiCoTeV} (+9.53\%) provides substantial contribution. The result indicate that the dynamic mutual validation scoring is a highly effective mechanism for enhancing reliability.

When evaluating two-module combinations, any pair of modules demonstrates a strong synergistic effect, with all combinations exceeding 90\% accuracy. 
More importantly, combining the high-quality test cases mined by \textbf{TCQM} with the subsequent modules yields substantial performance gains. 
When \textbf{TCQM} is integrated with the iterative optimization process of \textbf{Parallel TDD}, the performance reaches \textbf{92.68\%} (+\textbf{12.58\%}). Furthermore, pairing the test cases from \textbf{TCQM} with the filtering process of \textbf{BiCoTeV} achieves the highest two-module accuracy at \textbf{93.90\%} (+\textbf{13.80\%}). These results clearly show that the high-quality test cases provided by \textbf{TCQM} act as a critical foundation, generating substantial benefits when combined with either downstream optimization or filtering processes.

The full \textbf{MineValiCoder} framework, integrating all three modules, achieves the highest performance at \textbf{96.34\%}, representing a substantial \textbf{+16.24\%} absolute improvement over the baseline. This result demonstrates that each module targets a unique part of the code generation pipeline. Specifically, \textbf{TCQM} ensures test case quality and diversity, \textbf{Parallel TDD} drives iterative code optimization, and \textbf{BiCoTeV} provides robust final selection. 
Their integration creates a mutually reinforcing system that significantly enhances code accuracy, even when applied to a strong backbone such as GPT-4.

\begin{table}[t]
  \centering
  \caption{Effect of Self-Validation on generated test case quality. 
  \textbf{w/o SV} and \textbf{with SV} denote the variants without and with Self-Validation, respectively. \textbf{Avg.} denotes the average number of generated test cases per test suite for each problem.  \textbf{Val.} (Validity) refers to the ratio of generated test cases that can pass the ground-truth code.}
  \label{tab:self-validation-ablation}
  \begin{tabular}{lccc}
    \toprule
    \textbf{Benchmark} & 
    \begin{tabular}{@{}c@{}}\textbf{w/o SV} \\ \textbf{(Avg. / Val.)}\end{tabular} & 
    \begin{tabular}{@{}c@{}}\textbf{with SV} \\ \textbf{(Avg. / Val.)}\end{tabular} & 
    \begin{tabular}{@{}c@{}}\textbf{$\Delta$} \\ \textbf{(Avg. / Val.)}\end{tabular} \\
    \midrule
    HumanEval & 5.91 / 80.65\% & 5.30 / \textbf{93.34\%} & -0.61 / \textbf{+12.69\%} \\
    MBPP      & 5.02 / 82.61\% & 4.39 / \textbf{87.26\%} & -0.63 / \textbf{+4.65\%}  \\
    APPS      & 4.59 / 72.87\% & 4.02 / \textbf{84.11\%} & -0.57 / \textbf{+11.24\%} \\
    \bottomrule
  \end{tabular}
\end{table}

\paragraph{Ablation study of the Self-Validation}
To further investigate the effectiveness of the self-validation mechanism in improving generated test case quality, we conduct an additional ablation study across HumanEval, MBPP, and APPS. 
To ensure consistency with our main experiments—where Qwen3-4B is employed to generate test cases for the subsequent code optimization workflow—this ablation study also adopts Qwen3-4B as the backbone model. 
Notably, because LiveCodeBench lacks ground-truth code from dataset for its problems, it is excluded from this analysis.
Each entry in Table~\ref{tab:self-validation-ablation} is composed of two metrics: average (\textbf{Avg.}) and validity (\textbf{Val.}). 
Test case validity quantifies whether generated test cases can provide reliable execution feedback. 

Across all three benchmarks, self-validation improves test case validity while reducing the average number of retained tests. On HumanEval, validity increases from 80.65\% to 93.34\% (+12.69\%); on MBPP, it increases from 82.61\% to 87.26\% (+4.65\%); and on APPS, it increases from 72.87\% to 84.11\% (+11.24\%). These results show that self-validation improves the reliability of the test cases used by Parallel TDD and BiCoTeV by filtering erroneous tests.

\paragraph{Ablation study of the consistency-based robust filtering}

\begin{table}[t]
\centering
\caption{Ablation study of consistency-based robust filtering in BiCoTeV across different benchmarks.}
\label{tab:robust-filter-ablation-all}
\begin{tabular}{lccc}
\toprule
\textbf{Dataset} & \textbf{\begin{tabular}[c]{@{}c@{}}w/o Robust\\ Filtering (\%)\end{tabular}} & \textbf{\begin{tabular}[c]{@{}c@{}}with Robust\\ Filtering (\%)\end{tabular}} & \textbf{$\Delta$} \\
\midrule
HumanEval      & 92.10 & \textbf{96.34} & \textbf{+4.24} \\
MBPP           & 87.00 & \textbf{87.40} & \textbf{+0.40} \\
APPS           & 59.33 & \textbf{64.00} & \textbf{+4.67} \\
LiveCodeBench  & 47.33 & \textbf{51.33} & \textbf{+4.00} \\
\bottomrule
\end{tabular}
\end{table}

We conduct a comprehensive ablation study on four benchmarks using GPT-4 to explore whether consistency-based robust filtering alleviates overfitting to simple test cases and improves the reliability of final code selection.

As shown in Table~\ref{tab:robust-filter-ablation-all}, integrating robust filtering consistently yields performance gains across all datasets. Specifically, it improves the Pass@1 accuracy by +4.24\% on HumanEval, +0.40\% on MBPP, +4.67\% on APPS and +4.00\% on LiveCodeBench. These consistent improvements across diverse benchmarks confirm that our method effectively removes overfitted code candidates, reduces the randomness of LLM outputs, and enables trustworthy and robust code selection.

\subsection{Parameter Sensitivity and Cost-Efficiency Analysis}
\begin{figure}[h!]
    \centering
    \includegraphics[width=1.0\columnwidth]{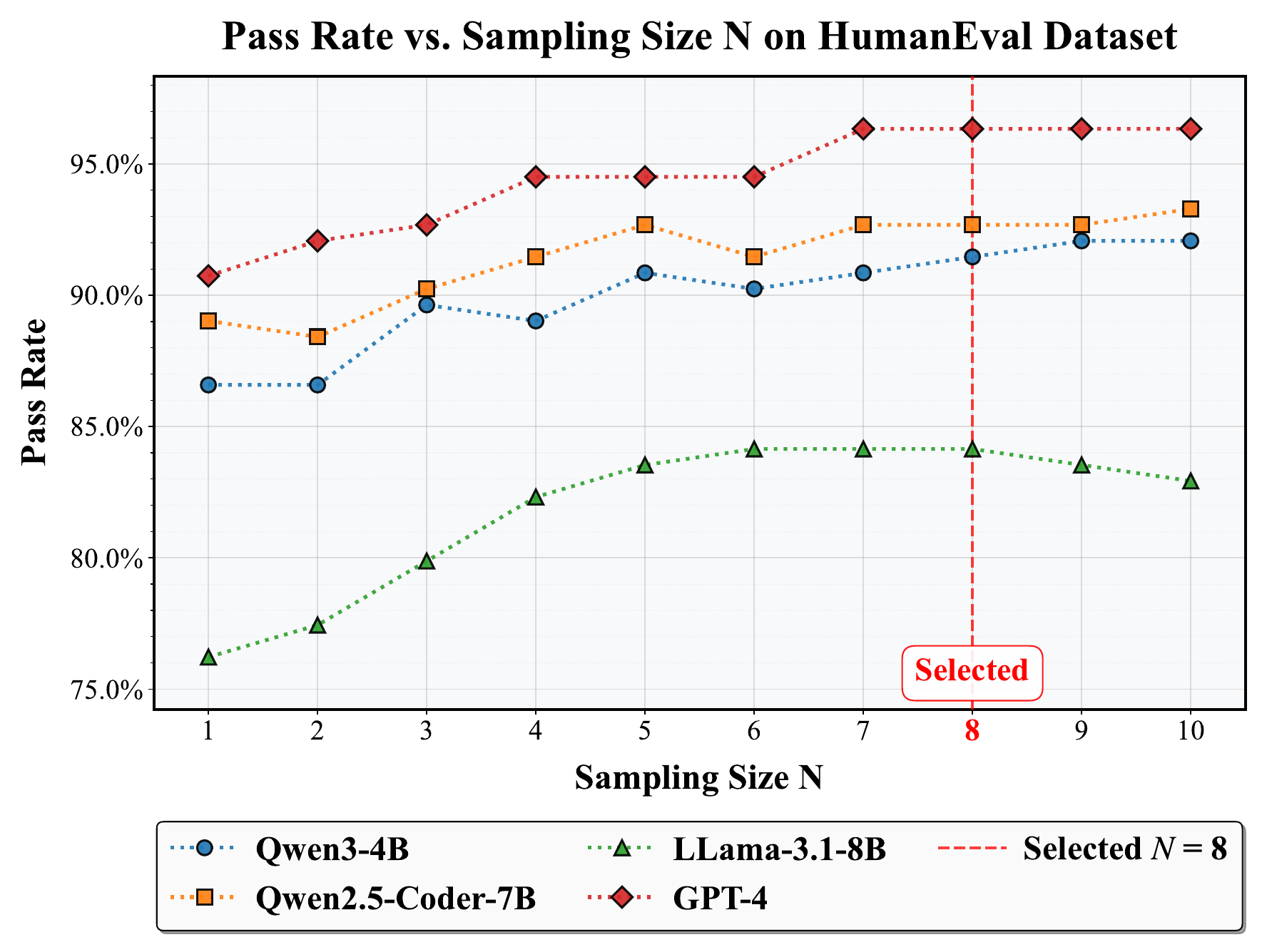}
    \caption{Performance sensitivity to the sampling size $N$ on HumanEval. The red annotation indicates our chosen setting ($N=8$), which offers the best trade-off between accuracy and cost.}
    \label{fig:impact_n}
\end{figure}

\paragraph{Impact of Sampling Size $N$}

To investigate the trade-off between performance gain and computational cost, we evaluated the Pass@1 accuracy on HumanEval across different backbones and sampling sizes $N$.
As a core hyperparameter of MineValiCoder, $N$ determines the number of TCQM generation rounds, Parallel TDD pipelines, and code candidates input to BiCoTeV.
Intuitively, a larger $N$ increases the probability of covering corner test cases during the TCQM module and provides a broader candidate pool for BiCoTeV module.
In our analysis, we set $N \in \{1, 2, \dots, 10\}$.

As illustrated in Figure~\ref{fig:impact_n}, the performance follows similar logarithmic growth trends across most models. 
For GPT-4, Qwen3-4B, and Qwen2.5-Coder-7B, increasing $N$ consistently improves accuracy, with three showing steady gains up to $N=8$. 
In contrast, Llama3.1-8B exhibits a different pattern: its accuracy peaks earlier and even slightly declines at higher $N$ values.
Our selected setting of $N=8$ achieves a balance between high reliability and inference efficiency across diverse model scales: GPT-4 (96.34\%), Qwen2.5-Coder-7B (93.29\%), Qwen3-4B (92.07\%), and Llama3.1-8B (84.76\%).
Although scaling to $N=10$ can push accuracy slightly further for some models, the marginal gains diminish while computational cost increases linearly. 
This linear growth occurs because each increment in $N$ necessitates an additional independent round for the Parallel TDD pipeline.

\begin{table}[t]
  \centering
  \caption{Comparison of average token consumption per problem and Pass@1 accuracy between MapCoder and MineValiCoder using GPT-4. Lower tokens and higher accuracy indicate better efficiency and performance.}
  \label{tab:cost-efficiency}
  \begin{tabular}{lcc}
    \toprule
    \textbf{Benchmark} & 
    \begin{tabular}{@{}c@{}}\textbf{MapCoder} \\ \textbf{(Tokens / Pass@1)}\end{tabular} & 
    \begin{tabular}{@{}c@{}}\textbf{MineValiCoder} \\ \textbf{(Tokens / Pass@1)}\end{tabular} \\
    \midrule
    HumanEval & 12.75k / 93.90\% & \textbf{3.95k} / \textbf{96.34\%} \\
    MBPP      & 4.96k / 83.10\%  & \textbf{3.89k} / \textbf{87.40\%} \\
    APPS      & \textbf{31.80k} / 46.67\% & 32.11k / \textbf{64.00\%} \\
    \bottomrule
  \end{tabular}
\end{table}

\paragraph{Cost and Efficiency Analysis}
We analyze average token consumption per problem to assess the computational cost of MineValiCoder relative to MapCoder, a representative multi-agent baseline with complex reasoning workflows. 
To clearly illustrate how token consumption scales with task complexity, we specifically conduct this analysis across three canonical benchmarks: HumanEval, MBPP, and APPS. 
These datasets are purposefully selected as they constitute a clear and progressive difficulty spectrum-ranging from fundamental functional programming to complex competition-level algorithmic tasks. 
While LiveCodeBench is utilized in our main experiments to evaluate temporal generalization against data contamination, its internally mixed difficulty levels make it less suited for analyzing the direct correlation between task complexity and computational cost. 

As shown in Table~\ref{tab:cost-efficiency}, MineValiCoder consumes significantly fewer tokens than MapCoder on HumanEval and MBPP. On the challenging APPS dataset, our token consumption remains comparable to MapCoder, yet MineValiCoder achieves a substantial absolute accuracy improvement of +17.33\% (64.00\% vs. 46.67\%). This high efficiency is primarily attributed to the concise design of our framework. 
By avoiding overly complex and redundant multi-agent interactions, MineValiCoder effectively minimizes unnecessary token generation while achieving superior accuracy.

\section{Conclusion}
In this paper, we introduced MineValiCoder, a fully automated code generation framework that integrates test case quality mining with bipartite graph-based mutual validation. 
MineValiCoder effectively addresses the challenges of invalid TDD feedback and result instability. 
Specifically, MineValiCoder integrates the Test Case Quality Mining (TCQM) module for high-quality test cases generation and employs a Parallel TDD module for stable code optimization. 
Finally, it utilizes the Bipartite Graph-Based Code-Test Mutual Validation (BiCoTeV) module for selecting the optimal code.
Extensive experiments across various models and benchmarks demonstrated that MineValiCoder achieves superior performance compared to state-of-the-art methods. 
Our findings highlight the effectiveness of autonomous test case mining and bipartite graph-based mutual validation in code generation. 
This suggests that MineValiCoder has the potential to empower further improvements in automated software engineering.

\bibliographystyle{IEEEtranN}
\bibliography{refs} 

\clearpage

\end{document}